\documentclass[nofootinbib,letter]{revtex4}

\usepackage{epsfig}
\newcommand{\beq}{\begin{equation}}
\newcommand{\eeq}{\end{equation}}
\newcommand{\beqa}{\begin{eqnarray}}
\newcommand{\eeqa}{\end{eqnarray}}

\newcommand{\ket}[1]{\ensuremath{{|{#1}\rangle}}}

\newcommand{\average}[1]{\ensuremath{{\langle{#1}\rangle}}}

\begin{document}

\title{`Flat Phase' Loading of a Bose-Einstein Condensate into an
Optical Lattice}

\author{Shlomo E. Sklarz}
\author{Inbal Friedler}
\author{David J. Tannor}
\affiliation{Department of Chemical Physics, Weizmann Institute of
Science, Rehovot, Israel 76100. Tel 972-8-9343723, Fax 972-8-9344123}
\author{Yehuda B. Band}
\affiliation{Department of Chemistry,
Ben-Gurion University of the Negev, Beer-Sheva, Israel 84105}
\author{Carl J. Williams}
\affiliation{Atomic Physics Division, {N}ational {I}nstitute of
{S}tandards and {T}echnology, 100 Bureau Drive, Gaithersburg,
Maryland 20899-8423, USA}

\begin{abstract}
It has been proposed that the adiabatic loading of a Bose-Einstein
Condensate (BEC) into an optical lattice via the Mott-insulator
transition can be used to initialize a quantum computer [D.~Jaksch,
{\it et al.}, Phys.~Rev.~Lett.~{\bf 81}, 3108 (1998)].  The loading of
a BEC into the lattice without causing band excitation is readily
achievable; however, unless one switches on an optical lattice very
slowly, the optical lattice causes a phase to accumulate across the
condensate.  We show analytically and numerically that a cancellation
of this effect is possible by adjusting the harmonic trap
force-constant of the magnetic trap appropriately, thereby
facilitating quick loading of an optical lattice for quantum computing
purposes.  A simple analytical theory is developed for a
non-stationary BEC in a harmonic trap.

\end{abstract}
\maketitle


\section{\label{sec:Intro}Introduction}

Experimental advances in manipulating and controlling
Bose-Einstein Condensates (BECs) of dilute atomic gases has
resulted in a remarkable series of experiments \cite{Anglin}.  One
theoretical proposal for quantum computing using atoms as qubits
is to first load the atoms that are in a BEC into an optical
lattice.  Then, by varying the intensity of a laser used to form
an optical lattice the BEC will undergo a quantum phase transition
from its BEC-like superfluid state to a Mott-insulator state
\cite{Jaksch}. This has recently led to a seminal experiment by
Bloch and collaborators \cite {Greiner}.

In principle, starting with a BEC in a trap and turning on an
optical lattice of sufficient well depth in a sufficiently
adiabatic manner will prepare the Mott-insulator state.  In
practice, it is easy to turn on the optical lattice adiabatically
with respect to band excitation (excitation from one band to
another); however, it is substantially more difficult to turn on
the optical lattice adiabatically with respect to quasi-momentum
excitation.  The second, more stringent form of adiabaticity
requires that the optical lattice be switched on slowly with
respect to mean-field interactions and tunneling dynamics between
optical lattice sites, and hence typically requires milliseconds
\cite{Band_02}.  We will refer to the first form of adiabaticity
as `{\em inter}band adiabaticity' and the second form as `{\em
intra}band adiabaticity'.  The intraband adiabaticity condition
has been demonstrated in one-dimensional lattices by Orzel {\it et
al.} \cite{Orzel} and ultimately led to the pioneering
experimental demonstration of the Mott-insulator transition
\cite{Greiner}.  When not otherwise specified, the terms {\it
adiabatic} and {\it nonadiabatic} in this article will refer to
{\em intra}band adiabaticity.

The goal of the present paper is to present a simple strategy for
remaining in the adiabatic regime while switching on the optical
lattice much faster than the millisecond time scales ordinarily
required for intraband adiabaticity.  The strategy is to
counterbalance the switching on of the optical lattice with an
appropriate change in the force constant of the trap.  This strategy
is shown to correct and prevent much of the quasi-momentum excitation
and resulting phase damage that arises from the nonadiabatic nature of
the switching.

More specifically, the switching on of an optical lattice potential
can divide a BEC into many individual pieces where phase coherence is
maintained across the whole condensate.  This phase coherence can be
seen by instantaneously dropping the lattice and looking at the
momentum distribution through time of flight measurements.  However,
because of a spatially dependent change in the density and thus the
mean-field per well site, one can end up with a quadratic phase
dependence developing along the lattice direction if one does not load
the lattice adiabatically with respect to quasi-momentum excitations
\cite{Band_02}.  Elsewhere \cite{Sklarz02.1} it has been shown, using
optimal control methods, that one can control the phase evolution to
obtain a flat phase at some final time by time varying the harmonic
trap force-constant of a confining external (typically magnetic) trap.
Here we show analytically and numerically that a complete cancellation
of the phase development is possible by appropriately adjusting the
external trap.

This paper will focus solely on one-dimensional lattices,
considering only the dynamics of the BEC along the lattice, and
will ignore effects transverse to the lattice.  It should be noted
that the effects of transverse excitation will show-up on time
scales inversely proportional to $\omega_{\perp}$ the transverse
trapping frequency which is typically long compared to the times
in the present paper.  Work is now in progress toward further
extending these results to two- and three-dimensions.  It is
expected \cite{BTM_02} that the squeezing of the BEC into the
transverse directions can also be treated using the above method,
namely by an appropriate adjustment of the trap in those
directions.

There have been a number of recent publications of both experimental
\cite{Pedri_01, Arimondo_01, Denschlag_02} and theoretical
\cite{Stringari_02, BTM_02} studies involving the loading of BECs in
one-dimensional lattices, and the resulting dynamics.  This paper is
related to these publications but focuses explicitly on a means of
quickly loading an optical lattice from a BEC for quantum computing
purposes, as well as for improving experimental signal to noise in
short time experimental studies of BECs.  Note that we consider the
regime where the density of the condensate is sufficiently large that
mean-field effects are not entirely negligible.  Experiments can be
carried out in the truly dilute gas regime where mean-field effects
are negligible \cite{Denschlag_02}.  However, reducing the condensate
density to such low values would have to be carried out adiabatically,
adversely affecting the time to load the optical lattice from the
initial (dense) BEC.

The outline of the paper is as follows: in section
\ref{sec:Problem} we define the problem.  In Sec.~\ref{sec:first},
a simple analytical theory is developed for a nonstationary 1D BEC
in a harmonic trap.  It is shown that a change in the density of
the condensate induces a time-varying phase across the condensate
that can be eliminated by a change in the harmonic force constant
of the trap. In section \ref{sec:then} it is shown that the effect
of switching on the optical lattice is to generate a new effective
normalization of the BEC and an analytical expression is obtained
for the modified harmonic trap force-constant that compensates for
the new effective normalization.  The analytical theory is in
excellent agreement with numerical simulations.  A modified
version of the theory in the regime where the nonlinear
interaction is strong and hence the the width of the condensate
differs from well to well is developed in Sec.~\ref{sec:finally}.
Section \ref{sec:Conclusions} contains the conclusion.


\section{\label{sec:Problem}Description of Problem}

We consider a 1D BEC confined by a harmonic trap and governed by the
Gross-Pitaevskii equation \beq i\hbar {\partial \over \partial
t}\ket{\psi} = (\hat K+\hat V+N U_0 |\psi|^2)\ket{\psi}, \eeq where
$\hat K=-{\hbar^2\over 2m}{\partial^2 \over \partial x^2}$ is the
kinetic energy operator, $\hat V$ is the external potential energy
operator to be discussed shortly and $NU_0$ is the nonlinear atom-atom
interaction strength, $N$ being the number of atoms and $U_0=4\pi
a_{0}\hbar ^{2}/m$ is the atom-atom interaction strength that is
proportional to the $s$-wave scattering length $a_0$.  The BEC is
initially in the ground state of the trap potential and is therefore
stationary.  An optical lattice is then switched on, having the effect
of separating the BEC wave packet into a series of localized pieces. 
The potential energy operator therefore takes the form $\hat
V(x,t)={1\over 2}m\omega_{t}^2x^2+S(t)V_0\cos^2(kx)$, where
$\omega_{t}$ is the trap frequency - which may be time dependent, $k$
is the laser field wave number, $V_0$ is the lattice intensity and
$S(t)$ is the function that switches-on the laser for the optical
lattice and goes from $S = 0$ at the beginning of the ramp-on of the
optical potential to $S = 1$ at the end of the switching on time
$\delta t_s$.  In applications to quantum computing, one often wants
to create an optical lattice with one atom per lattice site which will
serve as quantum bits.  However, due to the nonlinearity of the
equations, the condensate wave function develops a phase that varies
from lattice site to lattice site when the optical lattice is not
turned on adiabatically \cite{Band_02}.  Such a wave function can be
represented by a superposition of quasi-momentum states, and a
superposition of quasi-momentum corresponds to a higher energy state
and thus cannot give rise to the Mott-insulator state.  The problem we
address is the elimination of this phase profile by adjusting the trap
strength.  In the coming section we analyze the evolution of BEC wave
functions in harmonic traps, and consider the effect of switching on
the optical lattice.  Finally, a closed form for the precise time
dependence of the trap strength that will insure a flat phase for the
wavefunction for all times after the optical potential is fully turned
on is derived.

First, however, we transform the NLSE to dimensionless units $t\to
t/t_0$, $x\to x/x_0$ and $\psi\to\sqrt{x_0}\psi$ where for convenience
we choose $t_0={mx_0^2 \over 2\hbar}$.  Performing these transformations
we end up with a dimensionless NLSE
\beq
i {\partial \over \partial t} \psi(x,t)=\left(-{1\over 4}{\partial^2
\over \partial x^2}+K(t)x^2+S(t)V\cos^2(kx)+U|\psi|^2\right)\psi,
\eeq
where the trap force-constant $K=\omega_{t}^2t_0^2$, the field intensity
$V=V_0{t_0/\hbar}$ and the nonlinear coefficient $U=NU_0t_0/x_0\hbar$,
such that all space, time and energy quantities are now expressed in
units of $x_0$, $t_0$ and $\hbar/t_0$ respectively. \footnote{ We do
not
determine, at this point, any specific choice of $x_0$.  Note however,
that choosing $x_0=\lambda/\pi$, the optical wave length, yields for
the energy units, $\hbar/t_0={\hbar^2 k^2\over 2m}\equiv E_r$ which
is just the recoil energy.}


\section{Analytical Theory} \label{sec:Analysis}

\subsection{Dynamics of a Thomas-Fermi BEC in an Harmonic trap}
\label{sec:first}

Consider a normalized Thomas-Fermi type BEC wave function in a 
harmonic potential of the form
\beq
\psi(x,t) = \left\{\begin{array}{lll} \sqrt{3\over 4w}\sqrt{1-{x^2\over
w^2}}e^{i(bx^2+c)}&\quad&(x/w)^2\leq1 \\
0&\quad&(x/w)^2>1 \end{array} \right.,
\eeq
where the width $w(t)$ and phase components $b(t)$ and $c(t)$ are all
assumed to be time dependent.  We wish to analytically describe the
evolution of this wave function in a harmonic trap with trap
force-constant $K$ (we first consider the case where $K$ is constant
in time, but the equations of motion for $w(t)$, $b(t)$ and $c(t)$
remain valid even if $K$ varies with time).  Inserting the above wave
function into the dimensionless NLSE, we obtain, by considering
separately the real and imaginary parts, two equations involving the
three parameters $w(t)$, $b(t)$ and $c(t)$.  The imaginary part yields
\beq
-{\dot{w}\over 2w}(1-{2x^2\over w^2\alpha^2})=-{b\over 2}(1-{2x^2\over
w^2\alpha^2})\quad\Rightarrow \dot{w} = wb
\label{eqFPwdot},
\eeq
where $\alpha\equiv\sqrt{1-{x^2\over w^2}}$, and from the real part we
get
\beqa
-\dot{b}x^2-\dot{c}&=&{1\over 4w^2\alpha^4}+b^2x^2+Kx^2+U{3\over
4w}(1-{x^2\over w^2})\nonumber\\
&\approx& {1\over 4w^2}(1+{2x^2 \over w^2})+(b^2+K)x^2+U{3\over
4w}(1-{x^2\over w^2}).
\eeqa
In going to the last line we expanded ${1\over \alpha^4}$ in a Taylor
series in $x/w$, truncating after the second order. Comparing separately
the coefficients of $x^0$ and $x^2$, we obtain the following two equations
of motion for $b(t)$ and $c(t)$:
\beqa
\dot{b}&=&-{1\over 2w^4}+{3U\over 4w^3}-(b^2+K) ,
\label{eqFPbdot} \\
\dot{c}&=& -{1\over 4w^2}- {3U\over 4w} .
\label{eqFPcdot}
\eeqa
Taking a time derivative of Eq.~(\ref{eqFPwdot}) and using
Eq.~(\ref{eqFPbdot}) we find
\beqa
\ddot{w}&=&\dot{b}w+b\dot{w}\nonumber\\
&=&-{1\over 2w^3}+{3U\over 4w^2}-Kw\nonumber\\
&\equiv&-{\partial\over \partial w}V_e(w)\label{eqFPwddot},
\eeqa
with the effective potential $V_e(w)$ defined as
\beq\label{eqFPdefV}
V_e(w)\equiv -{1\over 4w^2}+{3U\over 4w}+{1\over 2}Kw^2.
\eeq
The time evolution of the wave function width, $w$, can therefore be
easily determined by considering the form of the potential $V_e(w)$.
Furthermore, by defining
\beq
p\equiv wb,
\eeq
we can formulate the equations for the conjugate variables $w$ and $p$
as a Hamiltonian system of equations with $H(w,p)=p^2/2+V_e(w)$ such that
\beqa
\dot{w}&=&{\partial\over \partial p}H=p\\
\dot{p}&=&-{\partial\over \partial w}H=-{\partial\over \partial
w}V_e.
\eeqa

\begin{figure}[htb]
\fbox{\epsfig{figure=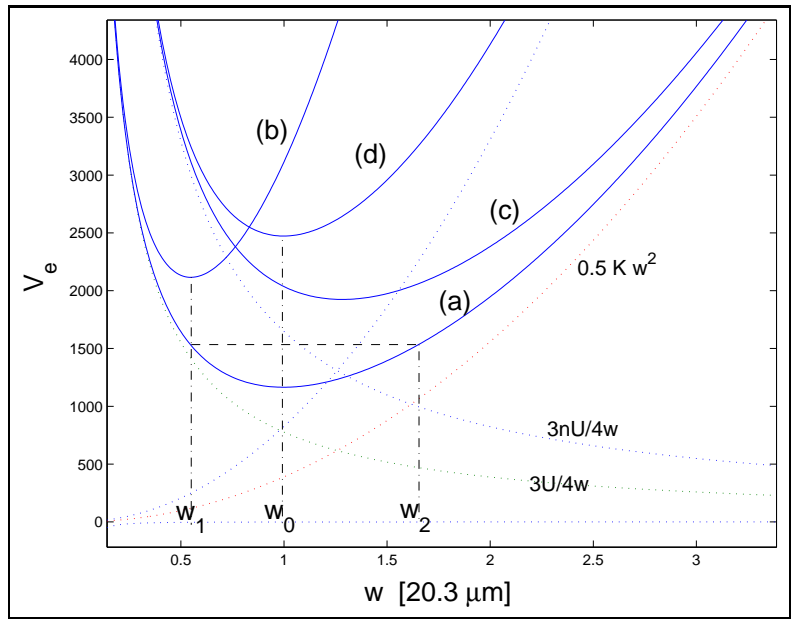,width=4in}} 
\caption{\label{VvsWpot}
a) $V_e(w)$ with stationary point $w_0$ and examples of turning
points $w_1$ and $w_2$ marked. The asymptotic curves correspond to the 
contributions of the two dominant terms in $V_e$ and highlight the 
way changes in these terms effect the dynamics. 
b) $V_e^{\prime}$ chosen such that
$w_0'=w_1$ by adjusting $K$.  c) $V_e^{\prime}$ obtained from (a)
as a consequence of change in wave function normalization.  d) a
new $V_e^{\prime}$ obtained from (c) by also changing $K$ to
compensate for the change affected by the normalization change
shown in (c). }
\end{figure}

Consider now the potential $V_e(w)$ in (\ref{eqFPdefV}) plotted as
curve (a) in Fig.~\ref{VvsWpot}.  The potential consists of a well
centered around the stable point $w_0 \approx ({3U\over
4K})^{1\over 3}$.  This can be most easily obtained by setting $b=0$
and $\dot{b}=0$ in Eq.~(\ref{eqFPbdot}) and solving for $w$ while
noticing that the first term on the RHS of (\ref{eqFPbdot}) is small
compared to the rest and can therefore be neglected.  With initial
wave function width $w(0)=w_0$, where $w_0$ is the width of the
Thomas-Fermi stationary ground state of the trap, the wave function
will remain stationary throughout.  However, if the initial width
equals some other value, an oscillatory motion of $w$ round the
stationary point $w_0$ will develop.  The phase curvature $b$ will
also oscillate with $w$ obtaining its maximum value when $w(t)=w_0$
and vanishing when $w$ approaches its turning points, $w_1$ and $w_2$.

If an abrupt change in the trap force-constant can be made, $K\to K'$, at
the exact point in time when $b(t)=0$, i.e., when $w(t)$ is at one
of its turning points, e.g., $w=w_1$, then it is possible to
change the potential so as to freeze the flat phased wave function and
make it stationary.  This can be obtained by choosing $K' =
{3U\over 4w_1^3}$ such that $w_1$ is the stationary point of the new
potential $V_e^{\prime}$ (curve (b) in Fig.~\ref{VvsWpot}).

Another scenario to be considered is the following.  We begin with a
stationary flat phased wave function residing at the stationary point
$w_0$ of the potential.  Imagine now the hypothetical possibility of
abruptly changing the normalization of the BEC wave function from
unity to $n$.  This would be equivalent to a change in the potential
$V_e\to V_e^{\prime}$ affected by changing $U\to nU$.  It is obvious
that this change will shift the stationary point to some new value
$w_0'=({3nU\over 4K})^{1\over 3}$ (see curve (c) in Fig.~\ref{VvsWpot})
and that the wave function currently positioned at $w_0$ will no
longer be stationary under the new potential.  In order to compensate
for this change and keep the wave function stationary one can adjust
the trap force-constant and set $K'=nK$ such that ratio $U/K$ remains
constant and the stationary point $w_0'=w_0$ will not shift (see curve
(d) in Fig.~\ref{VvsWpot}).

We show in the following section that turning on an optical
lattice corresponds to a change in the normalization of the wave
function, so that the above scenario corresponds precisely to our
goal of achieving a flat phased BEC loading of an optical lattice.
It should be noted that the above analysis ignores gravity which
can be assumed to be orthogonal to the lattice direction. However,
even if gravity is along the lattice direction a similar analysis
holds but requires an additional linear offset.


\subsection{Switching on the Optical Lattice}
\label{sec:then}

\begin{figure}[htb]
{\epsfig{figure=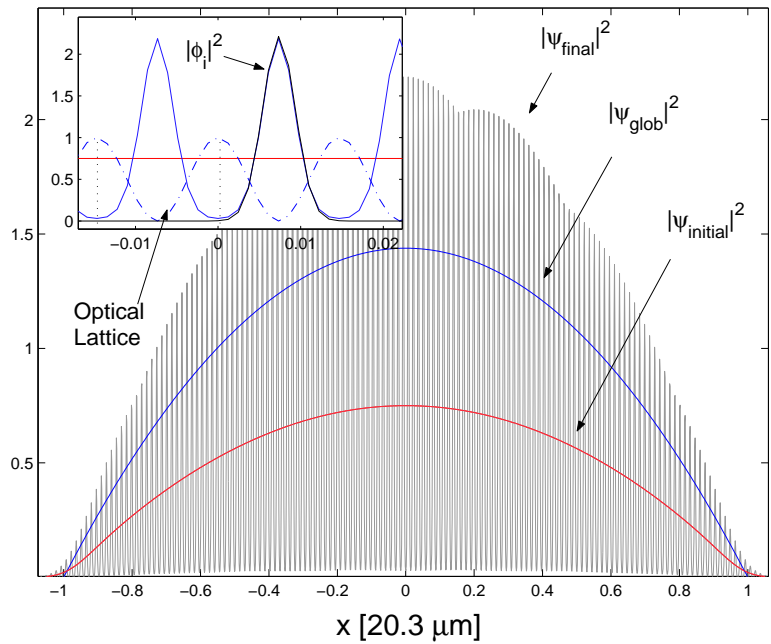,width=4in}}
\caption{\label{FltPhaseTheory} Analysis of the BEC wave function in
an optical lattice.  $\psi_{init}$ and $\psi_{final}$ are the wave
functions before and after applying the optical lattice, and
$\phi_i(x)$ is the local wave function within a specific well.
(Gaussian approximation) $\psi_{\mathrm{glob}}$ is the global
Thomas-Fermi type wave function after averaging out the local
details.}
\end{figure}

Quickly switching on the optical lattice causes the BEC wave function,
which initially has a Thomas-Fermi form of an inverted harmonic
potential, to split into a series of localized pieces each residing in
a lattice well.  As the overall normalization of the wave function
must remain unity, the displaced population from areas between the
lattice wells builds up within the wells such that the density in
these regions increases dramatically (see Fig.~\ref{FltPhaseTheory}). 
However, if we neglect the local lattice structure and consider solely
the global nature of the BEC wave function, we see that it retains its
quadratic shape, and the change in the wave function brought about by
the existence of the optical lattice can be viewed as a stretching of
the Thomas-Fermi wave function in the vertical direction (see
Fig.~\ref{FltPhaseTheory}).  This picture is based on a separation of
scales in the spatial dimension which is a consequence of the fact
that the length of each lattice well, ${\lambda\over 2} = {\pi\over
k}$, is much smaller than the scale of the total wave packet, $w$ (see
for example Ref.~\cite{Stringari_02}).  It is for this reason that we
can treat first the local structure of the wave function in each well
and then consider separately the overall global evolution of the wave
function.

The idea is therefore to view the wave function on a level coarser
than the lattice site dimension, averaging out the local lattice
structure of the wave function.  This procedure yields a new
Thomas-Fermi type wave function $|\psi_{\mathrm{glob}}|^2 =
\average{|\psi|^2}_{\mathrm{loc}} = {3n\over 4w}{(1-{x^2\over w^2})}$
differing from the initial one by a modified normalization factor $n$
(see Fig.~\ref{FltPhaseTheory}).  The evolution of this wave function
can then be analyzed using the results of the previous section.

This procedure can also be viewed as a spatial-averaging out of the
local structure of the Hamiltonian operator
$\average{H}_{\mathrm{loc}} = \average{T+V_{\mathrm{lattice}} + V_{t}
+ U|\psi|^2}_{\mathrm{loc}} = T^{\mathrm{av}} +
V^{\mathrm{av}}_{\mathrm{lattice}} + V_{t} +
U\average{|\psi|^2}_{\mathrm{loc}}$.
The harmonic trap potential $V_{t}$ is
constant on the local scale and is therefore unaffected by the
averaging.  If the average kinetic and lattice potential energies per
particle, $T^{\mathrm{av}}$ and $V^{\mathrm{av}}_{\mathrm{lattice}}$,
are constant from well to well, these contributions to the energy can
be absorbed into the chemical potential $\mu$, resulting in just the
averaged global mean-field playing-off, on the global scale, against
the trap potential as in a simple Thomas-Fermi procedure.  The trap
must then be adjusted to compensate only for the varying mean-field
across the BEC wave function.

In obtaining this simplified picture we distinguish between two
opposite scenarios occurring on the local scale.  In many cases, when
considering the dynamics along the direction of the one-dimensional
lattice, the mean-field within each well is negligible in comparison
with the kinetic and potential energies along this direction.  This
occurs for tight optical wells, e.g., short wavelength and strong
intensity such that $\sqrt{V}k^2\gg U/w$, where $w$ is the width of
the BEC. The local wave function can then be well approximated by a
Gaussian with a ``well-independent'' width implying that the locally
averaged kinetic and lattice potential energies are also
``well-independent''.  In carrying out the above procedure we then
find that the global wave function is a stretched image of the initial
one, as described above.

In the opposite regime the mean-field within each well can no longer
be neglected.  In these cases the calculations are more involved and
do not yield the simplified picture presented here of a mere
stretching of the wave function.  Instead, a distortion occurs which
must be treated explicitly.  We therefore delay discussion of this
scenario and provide a more general treatment in the next section.

In the following we wish to determine the normalization factor $n$ in
terms of the optical lattice parameters $V$ and $k$.  Consider the initial
Thomas-Fermi wave function $|\psi|^2={3\over 4w}(1-{x^2\over w^2})$.
The  number of atoms in the region of each lattice well determined
by its position $x_i$ is
\beqa\label{eqFPnloc}
\eta(x_i)&=&|\psi(x_i)|^2{\lambda\over 2}\nonumber\\
&=&{3\pi\over 4wk}(1-{x_i^2\over w^2}).
\eeqa

Assuming that the local population becomes trapped in the well during
the switching on of the optical lattice, we can then consider the
local normalization factor per well as constant throughout the
evolution.  Assuming too that the wave function at each lattice site
is localized after the optical lattice has been switched on, we can
ascribe to each lattice site a local wave function, $\phi_i(x)$, which
is normalized to $\eta_i$.  In order to obtain an average norm per
well we define the local probability function $P_i(x)$ which is just
the local wave function normalized to unity
\beq 
P_i(x)={1 \over \eta_i}|\phi_i(x)|^2.
\eeq
Averaging out the local structure using the local probability
function, $P_i(x)$, we obtain the coarse-grained wave function,
\beqa
\label{eqPsiglob}
|\psi_{\mathrm{glob}}(x_i)|^2=\average{|\phi_i|^2}_{\mathrm{loc}}
&=&\int P_i(x)|\phi_i(x)|^2 dx\nonumber\\
&=&{1\over \eta_i}\int|\phi_i(x)|^4dx.
\eeqa

Note, the limits of integration in the above integral should be
restricted to a single well but due to the gaussian-like nature of the
wavefunction $\phi_i(x)$ the specific limits are unimportant.  Note
that in evaluating the integral, $\eta_i$ was considered constant as
it is only slowly varying on the local scale.

In many cases the local wave function can be well approximated by
a Gaussian
\beq
\phi_i(x) =\sqrt{\eta_i\over \pi^{1/2}\Delta}
e^{-{(x-x_i)^2\over 2\Delta^2}}e^{i\Phi},
\eeq
where $\Delta$ is the width and the wave function normalizes to
the local normalization factor $\eta_i$ (see inset in
Fig.~\ref{FltPhaseTheory}).  $\Delta$ is of typically on the order
of but smaller than $\lambda\ll w$ and is therefore small compared
with the width of the total wave function, so $\eta(x_i)$ is only
slowly varying with respect to $x$ and can be considered constant
within any given lattice site. Averaging out the local structure
we obtain the coarse wave function $\psi_{\mathrm{glob}}(x_i)$
which we now show to be of Thomas-Fermi type
\beqa
|\psi_{\mathrm{glob}}(x_i)|^2&=&\average{|\phi_i|^2}_{\mathrm{loc}}=
        \int P_i(x)|\phi_i(x)|^2 dx \nonumber \\
&=& {\eta_i\over \sqrt{\pi}\Delta}{1 \over \sqrt{\pi}\Delta}
\int e^{-2{(x-x_i)^2\over \Delta^2}} dx \nonumber \\
&=&{\eta_i\over \sqrt{2\pi}\Delta} \nonumber \\
&=&{3\over 4w}\sqrt{\pi\over 2}{1\over k\Delta}(1-{x_i^2\over w^2})
\nonumber\\
&\equiv&{3n\over 4w}(1-{x_i^2\over w^2}).
\eeqa
In going from the third to the fourth line we used the explicit
form of $\eta_i$ given in (\ref{eqFPnloc}).  Comparing the last
two lines we find the modified normalization to be
\beq\label{eqFPn1} n=\sqrt{\pi\over 2}{1\over k\Delta }. \eeq It
remains to determine the local width $\Delta$ of the wave function
within each lattice site in terms of the external parameters.  It
can be shown analytically (see appendix \ref{variational}) that
\beq
\Delta=\sqrt{-{2\over k^2}\mathcal{W}(-{k\over 4 \sqrt{V}})}
\eeq
where $\mathcal{W}(x)$ is the Lambert W function \cite{LambertW}, so
that the normalization factor $n$ is finally given by
\beq
\label{eqFPn2} n={1\over 2}\sqrt{-\pi\over \mathcal{W}(-{k\over 4 
\sqrt{V}})}.
\eeq

The effect of switching on the optical lattice on the dynamics of the
wave function can now be viewed as changing the normalization of the
initial wave function from unity to $n$.

\begin{figure}[htb]
{\epsfig{figure=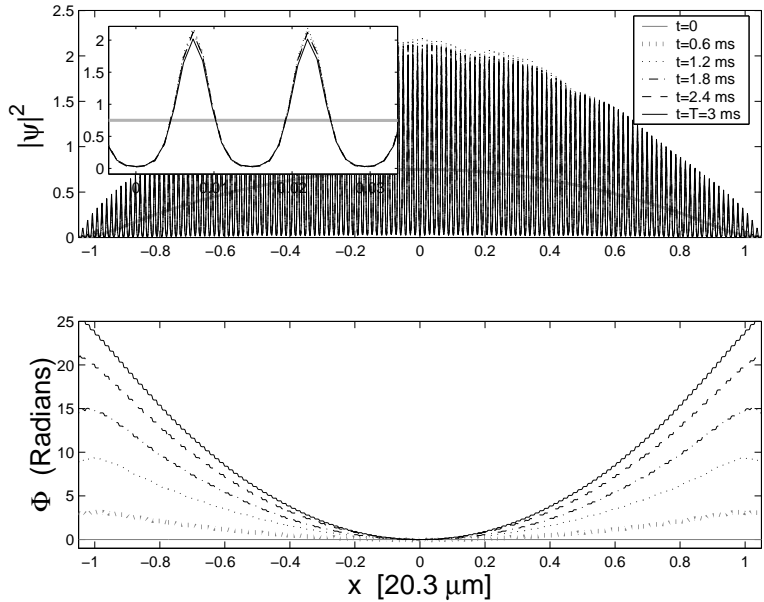,width=4in}}
\caption{\label{CurvdPhaseWF} Evolution of the wave function (amplitude and
phase) as a consequence of switching on the optical lattice. Note the
development of a quadratic phase profile. }
\end{figure}

\begin{figure}[htb]
{\epsfig{figure=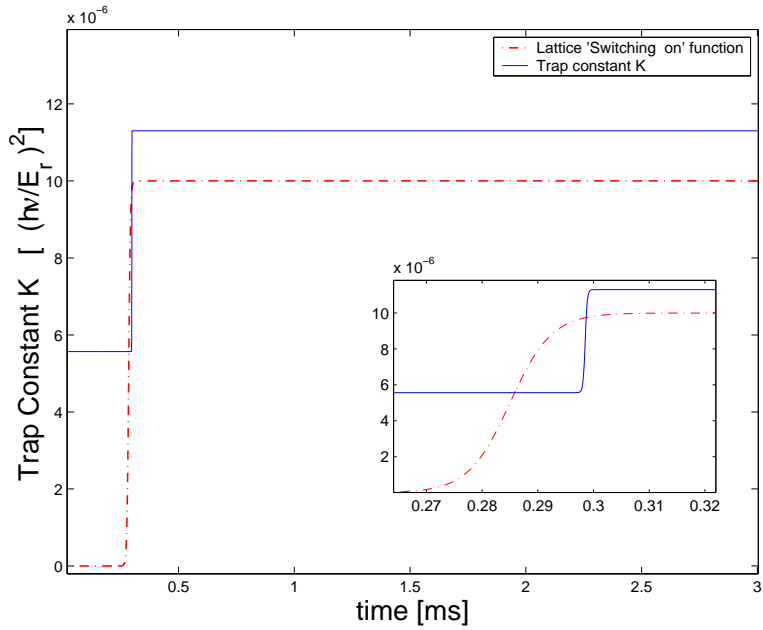,width=4in}}
\caption{\label{FltPhaseSeq} Sequence of external fields keeping phase
of wave function flat and stationary. }
\end{figure}

\begin{figure}[htb]
{\epsfig{figure=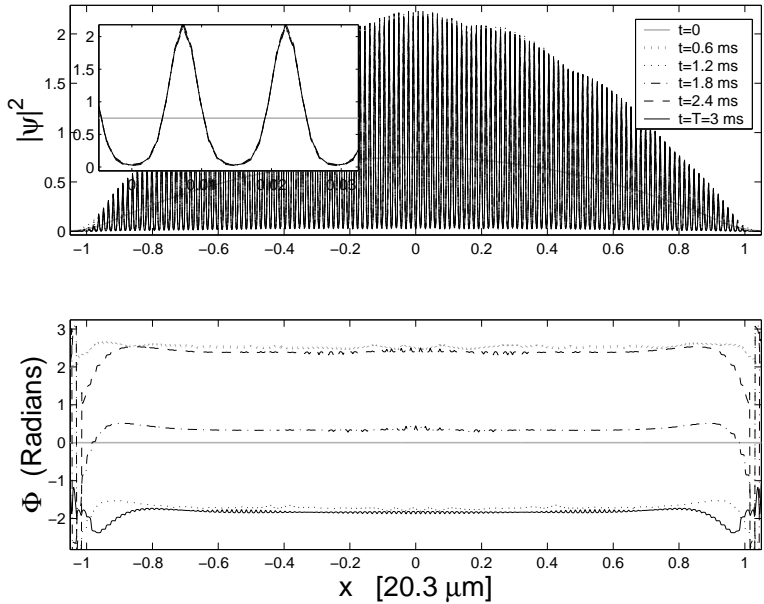,width=4in}}
\caption{\label{FltPhaseWF}
Evolution of  the wave function (amplitude and phase) as a consequence of
switching on the optical lattice and adjusting the compensating trap
force-constant.  The stationary flat phase is strikingly apparent. }
\end{figure}

If the switching-on time $\delta t_s$ is short compared to the global
nonlinear time scale $t_{NL}$ \cite{Trippenbach00} so as no
substantial phase evolution occurs during this time, the
transformation of the normalization constant can be considered abrupt
and the dynamics of the wave function parameters $w(t)$ and $b(t)$ are
raised from the initial potential curve $V_e$ to $V_e^{\prime}$
(curves (a) and (c) respectively in Fig.~\ref{VvsWpot}) by the change
$U\to nU$ as described in the previous section.  If no further
adjustments are made, the wave function will begin to evolve on the
potential curve $V_e^{\prime}$ and develop phase as seen in
Fig.~\ref{CurvdPhaseWF}.  In order to cancel this effect one can
compensate for the change of normalization by adjusting the trap
force-constant to $K=nK_0$ (curve (d) of Fig.~\ref{VvsWpot}).  In
Fig.~\ref{FltPhaseSeq} we plot the switching-on function of the
optical lattice and the change of the trap force-constant $\Delta K =
K-K_0$ as a function of time.  The evolution of the wave packet under
this sequence of events is plotted in Fig.~\ref{FltPhaseWF}, from
which it is evident that the phase remains constant throughout the
evolution for the correct tuning of the trap force-constant.

In the simulations presented here we have taken $N=1.5\times 10^6$
sodium atoms, a scattering length of $a_0=2.8$ nm, and a trap of
average frequency $59.26$ Hz.  Using these values the Thomas-Fermi
approximation to the chemical potential, $\mu_{\mathrm{TF}}$, can be
calculated and the nonlinear interaction time becomes
$t_{NL}\equiv{\hbar/\mu_{\mathrm{TF}}} = 96.2 \,\mu$s.  In order to
preserve the time scales in the 1D model as they are in 3D reality, we
follow Ref.~\cite{Trippenbach00} and replace the nonlinear coefficient
$NU_0$ by $C\mu_{\mathrm{TF}}x_{\mathrm{TF}}$, where the Thomas-Fermi
radius $x_{\mathrm{TF}} = \sqrt{{2\mu_{\mathrm{TF}}/m\omega^2_{t}}}$
gives the size of the condensate and the factor $C$ carries the
dependence of the simulation on the dimensions and is for our 1D case
$C={\sqrt{\pi}\over \Gamma(2+1/2)}={4\over 3}$ \cite{Trippenbach00}.

We take the optical lattice wavelength to be $\lambda=589$ nm and
choose $x_0=\lambda/\pi=2/k$ such that $t_0=\hbar/E_r$, where $E_r
\equiv {\hbar^2k^2\over 2m}$ is recoil energy.  The optical lattice is
switched on in a time $ \delta t_s \approx 20 \mu s$ to the final
intensity of $V_0=10.94 \, E_r$.  In units of $x_0=\lambda/\pi$, $t_0$
and $\hbar/t_0=E_r$, for space, time and energy quantities
respectively, we therefore get the following unitless values; $k=2$
for the optical wave number, $K=({\hbar\omega_{t}\over
E_r})^2=5.615\times 10^{-6}$ for the initial trap force-constant,
$V=10.94$ for the final field intensity and $U = {4\over 3}{t_0\over
t_{NL}}{x_{\mathrm{TF}}\over x_0}=9.55$ for the nonlinear interaction 
strength.

Inserting these values into Eq.~(\ref{eqFPn2}) yields the
normalization factor $n=2.0866$ such that the trap force-constant
which we analytically predict to yield an optimally flat phase is
$K=1.172\times 10^{-5}$ ($85.5$ Hz).  This value is off by merely
$2\%$ from the empirically found optimal value of $K=1.151\times
10^{-5}$ ($84.8$ Hz) which generates the evolution plotted in
Fig.~\ref{FltPhaseWF}.  Some small residual spatially varying phase 
structure remains.  This structure is due to incomplete interband 
adiabaticity and can be reduced by increasing the switching-on 
time $\delta t_s$.


\subsection{Nonlinear Regime}
\label{sec:finally}

We now return to the more complicated scenario where the local
wave function has spatially varying contributions from the
mean-field term. Various complications arise in this regime which
must be solved individually.  The main complication is due to the
fact that when the mean-field is locally important it affects the
width and shape of the local wave functions such that they differ
from well to well as shown in appendix \ref{variational}.  This
implies that the average kinetic and lattice potential energies
also vary from well to well, affecting the phase accumulation.

Assuming that the local wave function can still be approximated by
a Gaussian along the lattice direction (as is the case unless the
local mean-field is larger than the kinetic energy) we can use the
results of appendix \ref{variational} to obtain the {\it
well-dependent} width $\Delta(x_i)$.  This can be inserted back
into Eq.~(\ref{eqVarEn}) to obtain the total local energy as a sum
of its contributions: the kinetic energy $T$, the lattice
potential energy $E_{\mathrm{lattice}}$, the trap potential energy
$E_{t}$ and the mean-field energy $E_{mf}$.  The chemical
potential associated with a specific lattice site is
\cite{Dalfovo99}
\beqa
\mu &=& {1\over \eta}(T+E_{\mathrm{lattice}}+E_{t}+2E_{mf}) \nonumber\\
&=&{1\over 8\Delta_i^2}+{V\over
2}(1-e^{-k^2\Delta_i^2})+V_{t}(x_i)+{\eta_i U\over \sqrt{2\pi}\Delta_i}.
\eeqa
In order to keep the phase evolution constant from well to well one
must adjust $V_{t}(x_i)$ such that it cancels all other $x_i$
dependencies (originating in $\Delta(x_i)$ and $\eta(x_i)$) and thus
makes $\mu$ independent of $x_i$.
Another complication arises from the fact that the optical lattice
must be switched on adiabatically with respect to interband
excitations (as stressed in the introduction), e.g., the switching-on
time $\delta t_s$ in our dimensionless units must be longer than
${2\pi\over k\sqrt{V}}$.  This means that for experiments in which the
lattice wavelength is large, the lower bound on the switching-on time
becomes comparable to $t_{NL}$ and considerable phase evolution will
occur during this time.

To avoid the phase winding during the switching-on time one must make
the trap frequency change gradually so as to compensate for the
changing shape of the wave function at intermediate times.  We assume
as a zeroth order approximation that a transition of the magnetic trap
from its initial to its final form using the same switching-on
function as the optical lattice, will momentarily compensate for the
changing shape of the wave function.  The relevant parts of the
potential terms in the Hamiltonian will take the following form
$(1-S(t))V_{t}^{\mathrm{init}} + S(t) V_{t}^{\mathrm{final}} + S(t) V
\cos^2(kx)$.

\begin{figure}[htb]
{\epsfig{figure=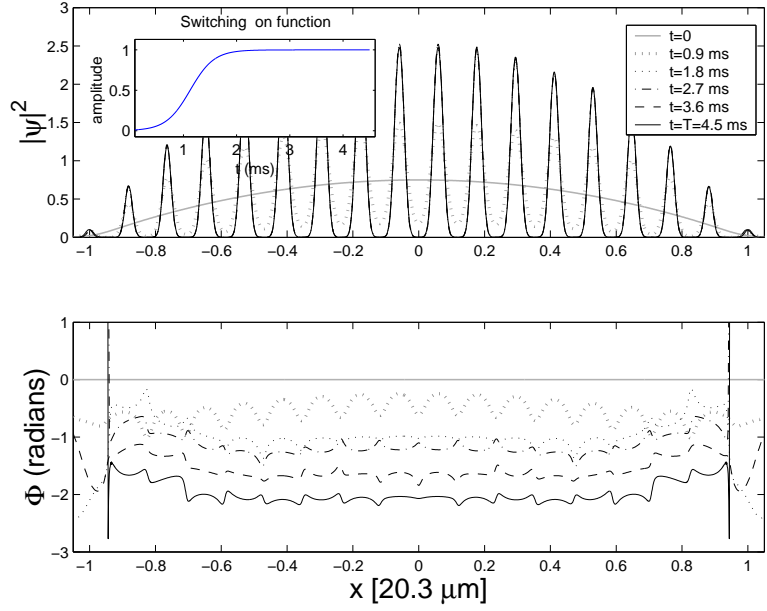,width=4in}}
\caption{\label{FltAdiaWF} Evolution of the wave function (amplitude and
phase) as a consequence of adiabatically switching on the optical
lattice and adjusting the compensating trap (strength and shape).}
\end{figure}

In Fig.~\ref{FltAdiaWF} we show a case where the mean-field is
important.  In this simulation we chose parameters as above, except
the optical lattice wavelength and strength were changed to be
$\lambda=8\times 589$ nm, \footnote{This can be accomplished by
changing the configuration of the lasers that make the optical lattice
from counter-propagating to intersecting at an angle $\theta$ such
that $\sin(\theta/2) = 1/8$ (i.e., $\theta = 14.36$ degrees).} and
$V_0=45.4\, E_r$ respectively so that the mean-field within each well
is no longer negligible.  With these parameters and following the
above procedure we found the optimal trap shape to be of the form
$V^{\mathrm{final}}_{t} = K\{1-(1-({x\over w})^2)^{0.8}\}$ with
$K=11.88$, where $w=x_{\mathrm{TF}}/x_0=13.5$ is the width of the BEC
in the units of $x_0$ introduced above.  We turned on the new trap
shape gradually, as described above, with a switching-on time of
$\delta t_s \approx 1$ ms, the resulting constant flat phase can be
clearly seen in Fig.~\ref{FltAdiaWF}.  Some small residual spatially
varying phase structure due to incomplete interband adiabaticity remains here
too, and increasing the switching-on time $\delta t_s$ will reduce the
residual phase structure.

\section{\label{sec:Conclusions}Conclusions}

The switching on of an optical lattice potential can divide a BEC into
many individual pieces where phase coherence is maintained across the
whole condensate.  However, because of a spatially dependent change in
the density and thus the mean-field per well site, one can end up with
a quadratic phase dependence developing along the lattice direction if
one does not load the lattice adiabatically.  We have shown
analytically and numerically that a cancellation of this effect is
possible by appropriately adjusting the external trap.  A simple
analytical theory has been developed for a non-stationary 1D BEC in a
harmonic trap.  It was shown that the effect of switching on the
optical lattice is to generate a new effective normalization of the
BEC, and hence a nonstationary condensate.  Finally, an analytical
expression was obtained for the modified harmonic trap force-constant
that compensates for the new effective normalization.  The analytical
theory is in excellent agreement with numerical simulations.

In real experiments more care is needed to account for the effects of
evolution in the transverse directions.  Work is now in progress
toward extending these results to two and three dimensions.  It is
expected that the expansion of the BEC into the transverse directions
can also be treated using the above method namely by an appropriate
adjustment of the trap in those directions.  We have detailed
elsewhere how our quasi-1D calculations of the type we presented here
model 3D aspects of the dynamics in cylindrically symmetric potentials
\cite{BTM_02}, but this method can not describe radial excitations of
the BEC that might arise due to the optical potential via the
mean-field interaction.  To the extent that radial excitations are not
important, our method should be an adequate approximation to the 3D
dynamics.

It is not known how a small residual spatially varying phase
will affect the Mott-insulator transition.
The residual phase can be thought of as a phonon-like excitation that 
should be mapped onto the
final Mott-Insulator state.
Characterizing the nature of excitations in an
inhomogeneous Mott-Insulator has not been done;
however, the small residual
excitations seen here are not expected to have a strong effect since the total
energy of the system is only slightly above that of the ideal case.
A more exact answer to this question 
can not be provided within the
context of a mean-field approach and requires analysis using many-body
approaches to the Mott-insulator transition. Moreover, no theoretical model
exists that is completely appropriate in both the superfluid and Mott-Insulator
regimes.


\appendix
\section{Calculation of local wave function width $\Delta$}
\label{variational}

As in the text we approximate the local wave function by a Gaussian
\beq
\phi_i(x) =\sqrt{\eta(x_i)\over \pi^{1/2}\Delta}e^{-{(x-x_i)^2 \over
2\Delta^2}},
\eeq
of width $\Delta$, normalized to the local normalization factor
$\eta_i$ and centered around $x_i$.  We wish here, using the
variational method, to determine the width of the Gaussian in terms of
the optical lattice strength $V$ and wave number $k$.

We first compute the energy associated with $\phi_i$ as a function
of $\Delta$:
\beqa\label{eqVarEn}
E(\Delta)&=&\int \phi_i^*(x)
                 \left(-{1\over 4}{\partial^2\over \partial
x^2}+V_{t}(x_i)+V\sin^2(k(x-x_i))+{1\over 2}U|\phi_i(x)|^2\right)
\phi_i(x) dx\nonumber\\
&=&({\eta(x_i)\over\sqrt{\pi}\Delta}) \int e^{-{(x-x_i)^2\over
2\Delta^2}} \left(-{1\over 4} {\partial^2\over \partial
x^2}+V_{t}(x_i)+V\sin^2(k(x-x_i))+{1\over 2}U|\phi_i
|^2\right)e^{-{(x-x_i)^2\over 2\Delta^2}}dx\nonumber\\
&=&\eta(x_i)\left({1\over8\Delta^2}+V_{t}(x_i)+{V\over
2}(1-e^{-k^2\Delta^2})+{\eta(x_i)U\over 2\sqrt{2\pi}\Delta}\right) .
\eeqa
Note that the trap potential, denoted $V_{t}$ and the number of
atoms in the region of the lattice well at position $x_i$,
$\eta(x_i)$, were extracted from the averaging integral since they are
assumed constant on the local scale.  According to the variational
principal, the determine the ground state function, $\phi_i$, which
minimizes the energy $E(\Delta)$ with respect to $\Delta$:
\beq\label{eqFPimplicitD}
{\partial E(\Delta)\over \partial \Delta} = \eta_i\left(-{1 \over
4\Delta^3} +V \Delta k^2e^{-k^2\Delta^2} - {\eta_iU\over 2\sqrt{2\pi}
\Delta} \right) = 0.
\eeq
An explicit solution of this equation for $\Delta$ is not possible in
general; we therefore distinguish between several cases and make some
simplifying assumptions. If, as is the case for short optical
wavelength, the mean-field term becomes negligible with respect to
the other energy terms, it can be neglected to obtain the following
equation
\beq
\Delta^4e^{-k^2\Delta^2}={1\over 4Vk^2}.
\eeq
The solution to this secular equation can be written in terms of the
Lambert $\mathcal{W}$ function, $y=\mathcal{W}(x)$, which is defined
as the inverse of $x=ye^y$ \cite{LambertW},
\beq
\Delta=\sqrt{-{2\over k^2}\mathcal{W}(-{k\over 4 \sqrt{V}})}.
\eeq
It can be seen in the inset of Fig.~\ref{FltPhaseTheory} that this
value for the width of the local wave function gives good results.  An
important point to note is that in this regime $\Delta$ is independent
of the well position, implying that the lattice potential and kinetic
energies per particle too are ``well-independent''.  This crucial
point justifies our treatment of the global wave function
$\psi_{\mathrm{glob}}$ as a Thomas-Fermi approximation.

For high density BECs and longer optical wavelengths the mean-field
cannot be neglected and Eq.~(\ref{eqFPimplicitD}) must be numerically
solved for $\Delta$. It must be noted, however, that the resulting
form for $\Delta(x_i)$ will in general be ``well-dependent'', implying
that the kinetic and lattice potential energies per particle will
also be ``well-dependent'' and thus contribute to the phase curvature
accumulation. This must be taken into account when adjusting the trap
to counter the phase accumulation, within the non-negligible mean-field
regime, as will be discussed in section \ref{sec:finally}.

\begin{acknowledgments}
This work was supported by 
the US Office of Naval Research (grant No. N00014-01-1-0667)
the Israel Science Foundation (grant No. 128/00-2),
and the German-israel BMBF (grant No. 13N 7947).
YB acknowledges support from the U.S.-Israel Binational Science
Foundation (grant No.~1998-421), the Israel Science Foundation (grant
No.~212/01) and the Israel Ministry of Defense Research and Technology
Unit.  CJW acknowledges partial support of the US Office of Naval
Research, the Advanced Research and Development Activity, and the
National Security Agency.

\end{acknowledgments}



\end{document}